\documentclass[%
reprint,
superscriptaddress,
showpacs,preprintnumbers,
amsmath,amssymb,
aps,
pra,
longbibliography
]{revtex4-2}
\usepackage{color}
\usepackage{graphicx}
\usepackage{bm}
\usepackage{mathrsfs}
\usepackage{dsfont}
\usepackage{ulem}
\usepackage[colorlinks,allcolors=blue]{hyperref}
\usepackage{subfigure}

\makeatletter

\newcommand{\Rmnum}[1]{\expandafter\@slowromancap\romannumeral #1@}

\makeatother

\begin{document}
\title{High-precision estimation of the parameters in the reservoir via the two-level system}

\author{Mengmeng Luo}
\affiliation{MOE Key Laboratory for Nonequilibrium Synthesis and Modulation of Condensed Matter, Shaanxi Province Key Laboratory of Quantum Information and Quantum Optoelectronic Devices, School of Physics, Xi'an Jiaotong University, 710049, P.R.China}

\author{Wenxiao Liu}
\affiliation{Department of Physics and Electronics, North China University of Water Resources and Electric Power, Zhengzhou, 450046, China}

\author{Yuetao Chen}
\affiliation{MOE Key Laboratory for Nonequilibrium Synthesis and Modulation of Condensed Matter, Shaanxi Province Key Laboratory of Quantum Information and Quantum Optoelectronic Devices, School of Physics, Xi'an Jiaotong University, 710049, P.R.China}

\author{Shaoyan Gao}
\email{gaosy@xjtu.edu.cn}
\affiliation{MOE Key Laboratory for Nonequilibrium Synthesis and Modulation of Condensed Matter, Shaanxi Province Key Laboratory of Quantum Information and Quantum Optoelectronic Devices, School of Physics, Xi'an Jiaotong University, 710049, P.R.China}
\date{\today}

\begin{abstract}
A scheme is proposed to estimate the system and environmental parameter, the detuning, temperature and the squeezing strength with a high precision by the two-level atom system. It hasn't been reported that the squeezing strength estimation through quantum Fisher information. We find entangled state and optimal superposition state are beneficial for parameter estimation with one-qubit probe by calculating quantum Fisher information and fidelity. And the fidelity between initial and final states of the atom can be improved via the two-qubit probe. Moreover, the phenomenon of quantum Fisher information return occurs when the detuning or the temperature is estimated. Our work provides a basis for precision measurement technology and quantum information processing.
\end{abstract}

\maketitle
\section{Introduction} % (fold)
\label{sec:introduction}
During the past decade, quantum parameter estimation has brought increased attention of researchers to the issues of quantum metrology and quantum information process. quantum Fisher information (QFI), used to evaluate the accuracy limits of quantum measurements, plays an important role in quantum parameter estimation. QFI has been widely researched in theoretical and experimental physics, such as  estimating the phase and frequency \cite{1,2,3,4,5,6,7,8,9,10,11,12,13,14,15,16,17,18,19}, quantifying quantum coherence \cite{20} and recognizing multiparticle entanglement \cite{21}. Besides, QFI is also applied in biology \cite{22}. It is universally accepted that entangled probe can achieve a better precision than unentangled ones \cite{23,24,25,26,27,28}. However, some research results indicate that not all entangled states are helpful for quantum metrology \cite{29,30}.

The detuning, temperature and  the squeezing strength  are  important parameters in physics. Consequently, improving the precision of the above parameters are one of the important subjects in quantum metrology. Some researchers have investigated the estimation of the detuning via QFI. Gammelmark et al. estimated the precision of the detuning, the Rabi frequency and the decay rate of the atom in open quantum systems \cite{31}. Kiilerich et al. showed the detuning estimation in a laser-driven $\Lambda$-type atom by multichannel photon counting \cite{32}. Dimani et al. found that optimized states compared with multiple single-photon states and NOON states could improve the precision of the detuning estimation over the standard quantum limit \cite{33}. More recently, Mogilevtsev demonstrated that the Heisenberg limit can be restored when the detuning is estimated in the $N$ two-level systems coupled with a single bosonic dephasing reservoir \cite{34}. But the realization of these schemes are quite complicated. 
Temperature, which is one of the seven fundamental physical quantities. At present, there have been many studies on improving the estimation precision of the temperature with quantum estimation theory \cite{35,36,37,38,39,40,41,42}. However, the model that the two-level atom is directly immersed in a thermal reservoir hasn't been utilized for the temperature estimation. Several researchers have reported that squeezed sate has the potential to improve the accuracy of phase estimation \cite{43,44,45,46,47}, but little attention has been given to estimate the squeezing strength.

The purpose of this paper is to improve the estimation accuracy of detuning, temperature and the squeezing strength via QFI. A simple model is established to calculate the QFI physical quantities mentioned above and the fidelity of the evolved state. Furthermore, we investigate the effects of system parameters on QFI and fidelity, the non-Markovian behavior, the relation between the QFI and fidelity. Besides, the study is extended to two-qubit probe. The results reveal entangled state and optimal superposition state are useful for the detuning, temperature and the squeezing strength estimations with one-qubit probe. And the fidelity is remarkably enhanced by replacing the one-qubit probe with two-qubit probe. This results will be applied in high-precision measurements of other physical quantities, the quantum thermometry and quantum information processing.

The rest of this paper is organized as follows. In section \uppercase\expandafter{\romannumeral2}, parameter estimation theory is overviewed simply in quantum system. In
section \uppercase\expandafter{\romannumeral3}, we introduce the model that the two-level systems are coupled to a Fock state  field, a thermal reservoir, and a squeezed vacuum reservoir, respectively. In section \uppercase\expandafter{\romannumeral4}, we show and analyze the QFI and fidelity based on various cases. Finally,
our conclusions are summarized in section \uppercase\expandafter{\romannumeral5}.

\section{Parameter estimation theory} % (fold)
\label{sec:Parameter estimation theory}
Considering the quantum estimation theory, the variation of a parameter $\phi$ can be estimated by the quantum Cram\'{e}r-Rao bound \cite{48,49}, $\delta \phi \geqslant 1/\sqrt{\nu F_{Q}}$, where $\nu$ is the number of experiments and $F_{Q}$ signify the QFI. It is obvious that a better precision can be obtained as the value of QFI increases.
Based on the density operator with spectrum decomposition under diagonalized representation, the density matrix $\rho$ can be written as
\begin{align}
\rho =\underset{i=1}{\overset{N}{\sum }}p _{i}\left\vert \psi
_{i}\right\rangle \left\langle \psi _{i}\right\vert, \tag{1}
\end{align}
where $p _{i}$ and $\psi _{i}$ are the eigenvalues and eigenvectors of $\rho$, respectively, The QFI can be expressed as follows

\begin{align}
F_{Q}(\phi )=\underset{i}{\sum }\frac{(\partial _{\phi }p _{i})^{2}}{%
	p _{i}}+2\underset{i\neq j}{\sum }\frac{(p _{i}-p _{j})^{2}}{%
	p _{i}+p _{j}}\left\vert \left\langle \phi _{i}|\partial _{\psi
}\psi _{j}\right\rangle \right\vert ^{2},\tag{2}
\end{align}
where the first and second parts include sums all except for  $p _{i}= 0$ and $p _{i}+p _{j}= 0$. respectively.

Fidelity is usually applied to measure the degree of similarity between two quantum sates \cite{50}. The fidelity of two quantum states is given by \cite{37}
\begin{align}
f(\rho _{0},\rho _{1})\!=\!\frac{1}{2}\{1\!+\!\overrightarrow{a}_{0}\!\cdot\!
\overrightarrow{a}_{1}\!+\![(1\!-\!\overrightarrow{a}_{0}\!\cdot\! \overrightarrow{a}%
_{0})(1\!-\!\overrightarrow{a}_{1}\!\cdot\!\overrightarrow{a}_{1})]^{1/2}\},\tag{3}
\end{align}
where $\overrightarrow{a}$ is the Bloch vector of density matrix $\rho$ under Bloch representation. Hence, the fidelity between the initial state and final state of the atom can be obtained. In this paper, it's worth noting that $\overrightarrow{a}_{0}$ and $\overrightarrow{a}_{1}$ are the Bloch vectors of the initial atomic state and the final atomic state, respectively.

\section{The model} % (fold)
\label{sec:The model}
In order to calculate the QFI of the detuning, temperature and  the squeezing strength, we assume the two-level systems interact with a Fock state field, a thermal reservoir and a squeezed vacuum reservoir, respectively. The atom-field interaction Hamiltonian can be written as Eq.~(4a) when the quality factor of the cavity is assumed as infinite, while the atom-reservoir interaction Hamiltonian is expressed as Eq.~(4b) \cite{51}
\begin{align}
H_{I}^{\prime }&=\hbar \lambda (\sigma _{+}\hat{a}e^{i\Delta t}+\hat{a}^{\dag }\sigma_{-} e^{-i\Delta t}),\tag{4a}\\
H_{I}^{\prime \prime }&=\hbar \sum_{k}g_{_{k}}(\sigma
_{+}\hat{b}_{k}e^{i(\omega_{0}-\nu_{k})t}+\hat{b}_{k}^{\dag }\sigma
_{-}e^{-i(\omega_{0}-\nu_{k})t}),\tag{4b}
\end{align}
where $\Delta$ ($\Delta=\omega_0-\omega$) means the detuning between an atom and a single mode cavity, $\nu_{k}$ is the frequency of $k$-th reservoir mode, $\sigma _{+}(\sigma_{-})$ indicates the rasing (lowing) operator of the atom, that is, $\left\vert e\right\rangle \left\langle g\right\vert $ ($\left\vert g\right\rangle \left\langle e\right\vert $), $\hat{a}^{\dag }(\hat{a})$ and $\hat{b}_{k}^{\dag }(\hat{b}_{k})$ denote the creation (annihilation) of cavity and reservoir, respectively. 
\subsection{one-qubit probe}
The initial state of the system is supposed as $\left\vert \psi _{(0)}\right\rangle=\cos (\alpha )\left\vert
e,n\right\rangle +\sin (\alpha )\left\vert g,n+1\right\rangle $ when the atom interacts with a Fock state field, and $n$ means the number of photons included in cavity. The initial state of the atom is prepared as $\left\vert \psi _{(0)}\right\rangle _{A}=\cos (\alpha )\left\vert
e\right\rangle +\sin (\alpha )\left\vert g\right\rangle $ when the atom interacts with the reservoir. The reange of $\alpha$ is from $0$ to $90^{\circ}$. The initial state are separable states when $\alpha=0^{\circ}$ or $90^{\circ}$, while the initial state is entangled state or optimal superposition state when $\alpha=45^{\circ}$.
\subsubsection{Fock state field}
To begin with, the atom is coupled to a cavity with a Fock state field. In the basis $\left\vert e,n\right\rangle$ and $\left\vert g,n+1\right\rangle$, using the schr\"{o}dinger equation and tracing over the cavity field degrees of freedom, the reduced density matrix is given by \cite{51}
\begin{align}
\rho_{1} (t)=\left(
\begin{array}{cc}
\rho _{11} & 0 \\
0 & \rho _{22}%
\end{array}%
\right),\tag{5}
\end{align}
where $\rho _{11}=\left\vert \beta _{1}\right\vert ^{2}$ and $\rho _{22}=\left\vert \beta _{2}\right\vert ^{2}$, with

\begin{align}
\beta _{1}&=e^{i\frac{\Delta }{2}t}\{\cos (\alpha )[\cos (\frac{1}{2}%
\widetilde{\Omega }_{n}t)-i\frac{\Delta }{\widetilde{\Omega }_{n}}\sin (%
\frac{1}{2}\widetilde{\Omega }_{n}t)]
\notag\\
&\quad-i\sin (\alpha )\frac{\Omega _{n}}{%
	\widetilde{\Omega }_{n}}\sin (\frac{1}{2}\widetilde{\Omega }_{n}t)\},\notag\\
\beta _{2}&=e^{-i\frac{\Delta }{2}t}\{\sin (\alpha )[\cos (\frac{1}{2}%
\widetilde{\Omega }_{n}t)+i\frac{\Delta }{\widetilde{\Omega }_{n}}\sin (%
\frac{1}{2}\widetilde{\Omega }_{n}t)]
\notag\\
&\quad-i\cos (\alpha )\frac{\Omega _{n}}{%
	\widetilde{\Omega }_{n}}\sin (\frac{1}{2}\widetilde{\Omega }_{n}t)\}, \tag{6}
\end{align}
here, $\widetilde{\Omega }_{n}=\sqrt{\Omega _{n}^{2}+\Delta ^{2}}$ and $\Omega _{n}=2\lambda \sqrt{n+1}$. Then we obtain the QFI of the detuning by substituting the eigenvalues and eigenvectors of $\rho_{1} (t)$ into Eq.~(2).

\subsubsection{the thermal reservoir}

For the purpose of estimating temperature with a high precision, we assume the atom interacts with a thermal reservoir. Then, the equation of reduced density matrix evolution of the atom is written as

\begin{align}
\frac{d\rho }{dt}&=-\frac{1}{2}\gamma (m+1)[\sigma _{+}\sigma _{-}\rho
-2\sigma _{-}\rho \sigma _{+}+\rho \sigma _{+}\sigma _{-}]\notag\\
&\quad-\frac{1}{2}\gamma
m[\sigma _{-}\sigma _{+}\rho -2\sigma _{+}\rho \sigma _{-}+\rho \sigma
_{-}\sigma _{+}], \tag{7}
\end{align}
where $\gamma$ is the atomic decay rate, and $m$ is the thermal average boson number, that is, $m=\frac{1}{e^{\frac{\hbar\nu_{k} }{k_{B}T}}-1}$.
The atomic density matrix elements are obtained as
\begin{align}
\rho _{11}&=\frac{m}{2m+1}+[\cos ^{2}(\alpha )-\frac{m}{2m+1}]e^{-\gamma
	(2m+1)t} ,\tag{8a}\\
\rho _{12}&=\cos (\alpha )\sin (\alpha )e^{-\gamma (m+\frac{1}{2})t},\tag{8b}\\
\rho _{21}&=\cos (\alpha )\sin (\alpha )e^{-\gamma (m+\frac{1}{2})t},\tag{8c}\\
\rho _{22}&=\frac{m+1}{2m+1}-[\cos ^{2}(\alpha )-\frac{m}{2m+1}]e^{-\gamma
	(2m+1)t}. \tag{8d}
\end{align}

Afterwards, the QFI of mean photon number can be given by Eq.~(2). Furthermore, the QFI of the temperature can be expressed by the QFI of the mean photon number for convenience of calculations
\begin{align}
F_{Q}(T)=F_{Q}(m)(\frac{\partial m}{\partial T})^{2},  \frac{\partial m}{\partial T}=\frac{\hbar \omega_{0} e^{\frac{%
			\hbar \omega_{0} }{k_{B}T}}}{(e^{\frac{\hbar \omega_{0} }{k_{B}T}}-1)^{2}k_{B}T^{2}}. \tag{9}
\end{align}
\subsubsection{the squeezed vacuum reservoir}

Now we assume the atom is immersed in a squeezed vacuum reservoir for estimating the precision of the squeezing strength. The equation of motion for the atomic density matrix is written as

\begin{align}
\frac{d\rho }{dt}&=-\frac{1}{2}\gamma (M+1)[\sigma _{+}\sigma _{-}\rho
-2\sigma _{-}\rho \sigma _{+}+\rho \sigma _{+}\sigma _{-}]\notag\\
&\quad-\frac{1}{2}\gamma
M[\sigma _{-}\sigma _{+}\rho -2\sigma _{+}\rho \sigma _{-}+\rho \sigma
_{-}\sigma _{+}]\notag\\
&\quad-\gamma e^{-i\theta} N\sigma _{-}\rho \sigma _{-}-\gamma
e^ {i\theta} N\sigma _{+}\rho \sigma _{+}, \tag{10}
\end{align}
where $M=\sinh ^{2}(r)$, $N=\cosh (r)\sinh (r)$, r is the squeezing strength. $\theta$ denotes the reference phase. For simplicity, the reference phase is fixed at $\theta=0$. The resulting equations are 
\begin{align}
\overset{\cdot }{\rho _{11}}&=\gamma M\rho _{22}-\gamma (M+1)\rho _{11},\tag{11a}\\
\overset{\cdot }{\rho _{12}}&=-\gamma (M+\frac{1}{2})\rho _{12}-\gamma N\rho
_{21},\tag{11b}\\
\overset{\cdot }{\rho _{21}}&=-\gamma (M+\frac{1}{2})\rho _{21}-\gamma N\rho
_{12},\tag{11c}\\
\overset{\cdot }{\rho _{22}}&=\gamma (M+1)\rho _{11}-\gamma M\rho _{22}. \tag{11d}
\end{align}

After some calculations, the atomic density matrix elements are given by
\begin{align}
\rho _{11}&=\frac{M}{2M+1}+[\cos ^{2}(\alpha )-\frac{M}{2M+1}]e^{-\gamma
	(2M+1)t},\tag{12a}\\
\rho _{12}&=\cos (\alpha )\sin (\alpha )e^{-\gamma (M+N+\frac{1}{2})t},\tag{12b}\\
\rho _{21}&=\cos (\alpha )\sin (\alpha )e^{-\gamma (M+N+\frac{1}{2})t},\tag{12c}\\
\rho _{22}&=\frac{M+1}{2M+1}-[\cos ^{2}(\alpha )-\frac{M}{2M+1}]e^{-\gamma
	(2M+1)t}.\tag{12d}
\end{align}

The QFI of the squeezing strength is obtained by plugging Eq.~(12) into Eq.~(2).

The fidelity of the atom in different cases can be worked out through putting the Bloch vector of the initial atomic density matrix and the finally atomic state of the above three cases into Eq.~(3), respectively.

Not that, the QFI of the estimated parameters are zero at $t=0$, therefore, the time will start at one for calculating the QFI.
\subsection{two-qubit probe}
Now the two-qubit probe is considered for the parameter estimation. The initially atomic state is prepared in the maximal entanglement state $\left\vert \psi _{(0)}\right\rangle _{S}=\frac{1}{\sqrt{2}} (\left\vert
e_{A},g_{B}\right\rangle +\left\vert g_{A},e_{B}\right\rangle) $. And the cavity field is initially prepared in Fock state $\rho_{f}(0)=\left\vert n\right\rangle \left\langle n\right\vert$. When two two-level atoms  interact with the Fock state field, the wave function can be written as
\begin{align}
\psi (t)&=C_{1}\left\vert e_{A},e_{B},n-1\right\rangle +C_{2}\left\vert
e_{A},g_{B},n\right\rangle \notag\\
&\quad+C_{3}\left\vert g_{A},e_{B},n\right\rangle
+C_{4}\left\vert g_{A},g_{B},n+1\right\rangle,\tag{13}
\end{align}
for $n=0$ and the neglect of the dipole-dipole interaction between two atoms, the probability amplitudes are given by
\begin{align}
C_{1}&=0,\tag{14a}\\
C_{2}&=\frac{\cos (\alpha )+\sin (\alpha )}{2}[\cos (\frac{\Omega t}{2})-\frac{%
	i\Delta }{\Omega }\sin( \frac{\Omega t}{2})]e^{i\frac{\Delta }{2}t}\notag\\
&\quad+\frac{%
	\cos (\alpha )-\sin (\alpha )}{2},\tag{14b}\\
C_{3}&=\frac{\cos (\alpha )+\sin (\alpha )}{2}[\cos (\frac{\Omega t}{2})-\frac{%
	i\Delta }{\Omega }\sin( \frac{\Omega t}{2})]e^{i\frac{\Delta }{2}t}\notag\\
&\quad+\frac{%
	\sin (\alpha )-\cos (\alpha )}{2},\tag{14c}\\
C_{4}&=-[\cos (\alpha )+\sin (\alpha )]\frac{2i\lambda }{\Omega }\sin (\frac{%
	\Omega t}{2})e^{-i\frac{\Delta }{2}t},\tag{14d}
\end{align}
here $\Omega=\sqrt{8\lambda^{2}+\Delta ^{2}}$, $\alpha=45^{\circ}$. After tracing over the field, the evolved state of the system can be obtained as
\begin{align}
\rho _{1}(t)=\left( 
\begin{array}{cccc}
0 & 0 & 0 & 0 \\ 
0 & \rho _{22} & \rho _{23} & 0 \\ 
0 & \rho _{32} & \rho _{33} & 0 \\ 
0 & 0 & 0 & \rho _{44}%
\end{array}%
\right),\tag{15}
\end{align}
where $\rho _{22}=\left\vert C_{2}\right\vert ^{2}$, $\rho _{23}=C_{2}C_{3}^{\ast
}$, $\rho _{32}=C_{3}C_{2}^{\ast }$, $\rho _{33}=\left\vert C_{3}\right\vert
^{2}$ and $\rho _{44}=\left\vert C_{4}\right\vert ^{2}$.

When the atoms interacts with the thermal reservoir or squeezed vacuum reservoir, the evolved state of the system  are shown as follows
\begin{align}
\rho _{2}(t)&=\left( 
\begin{array}{cccc}
\rho _{11} & 0 & 0 & 0 \\ 
0 & \rho _{22} & \rho _{23} & 0 \\ 
0 & \rho _{32} & \rho _{33} & 0 \\ 
0 & 0 & 0 & \rho _{44}%
\end{array}%
\right),\tag{16a}\\
\rho _{3}(t)&=\left( 
\begin{array}{cccc}
\rho _{11} & 0 & 0 & \rho _{14} \\ 
0 & \rho _{22} & \rho _{23} & 0 \\ 
0 & \rho _{32} & \rho _{33} & 0 \\ 
\rho _{41} & 0 & 0 & \rho _{44}%
\end{array}%
\right),\tag{16b}
\end{align}
where the elements of the reduced density matrix are expressed as Eqs.~(A2) and (B2) of Ref.~[52], respectively. And for simplicity, we assume $a=0$, $m=n$, $\gamma_{1}=\gamma_{2}$, $r_{1}=r_{2}$ and $\theta_{1}=\theta_{2}=0$. 

For two-qubit probe, the precision of the detuning, temperature and the squeezing strength are estimated by putting Eqs.~(15) and (16) into Eq.~(2), respectively.

After tracing over the subsystem B, the recuced density matrix of subsystem A of Eqs.~(15) and (16) are given by 
\begin{align}
\rho^{A}(t)=\left( 
\begin{array}{cc}
\rho _{11}+\rho _{22} &  0\\ 
0 & \rho _{33}+\rho _{44}%
\end{array}%
\right),\tag{17}
\end{align}
then the fidelity between initial and final states of the atom A is obtained with the two-qubit probe for different cases.
\section{Numerical results and discussion}
\label{sec:Numerical results and discussion}
The main purpose of this paper is to improve the estimation accuracy of the detuning, temperature and the squeezing strength. As a result, the effect of initial atomic state on the  evolution of QFI is investigated with one or two qubits probe.  In addition, the fidelities of the atom in different cases are studied.

\begin{figure}{}
	\centering
	\subfigure[]{
		\includegraphics[width=6cm]{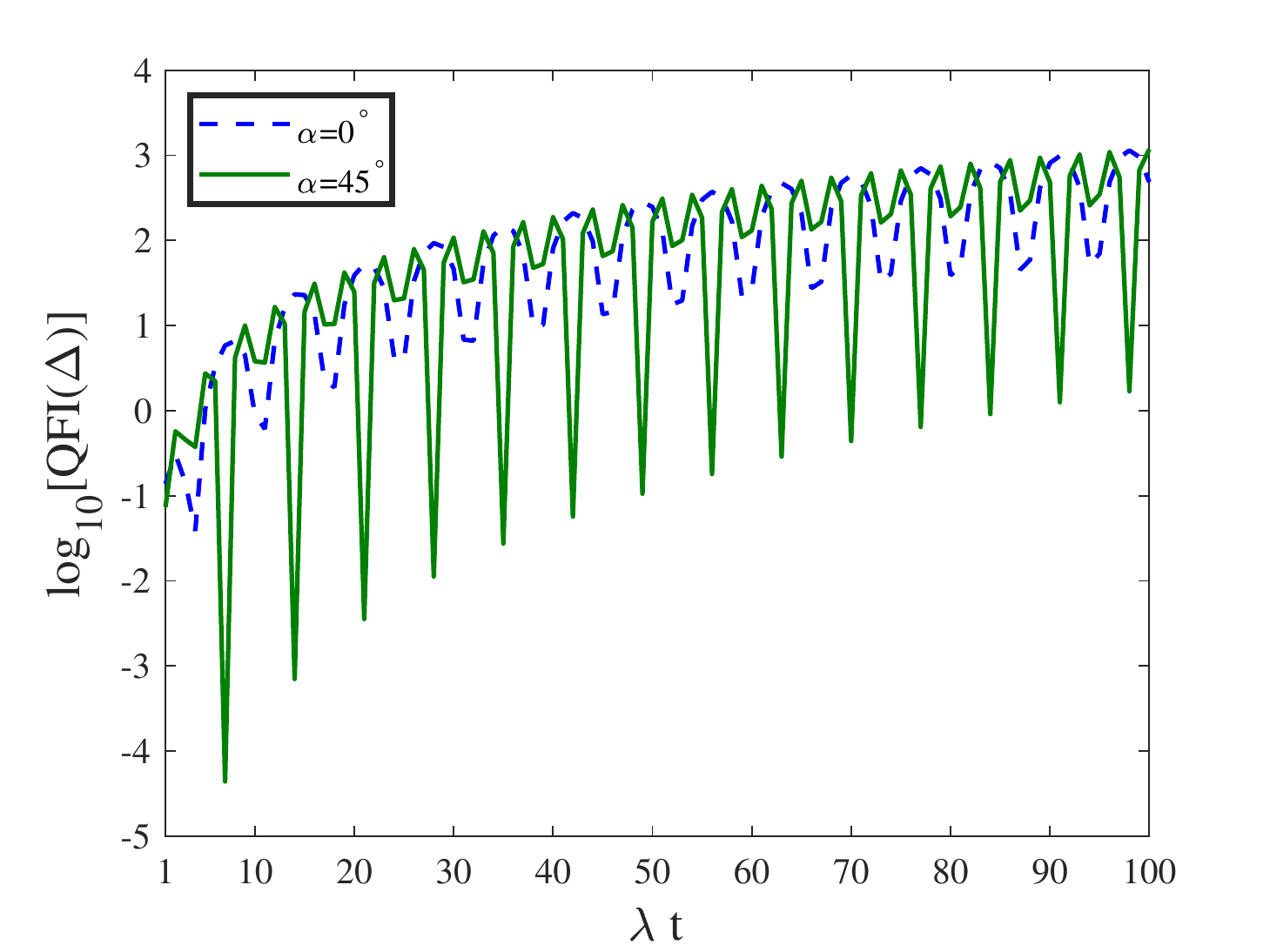}}
	\subfigure[]{
		\includegraphics[width=6cm]{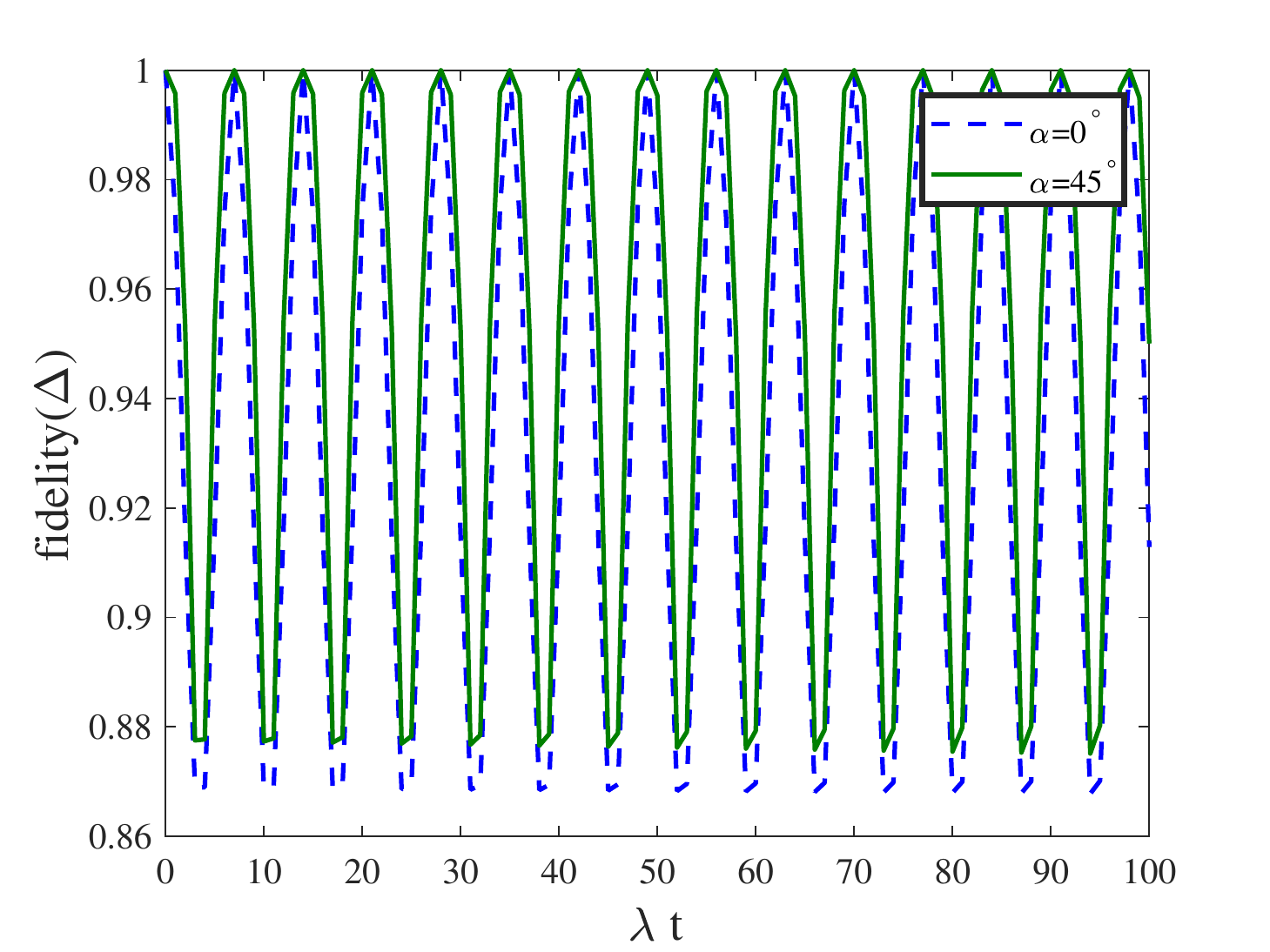}}
	\caption{\label{fig_double} (Color online) (a) QFI of the detuning $\Delta$ as a function of time for different $\alpha$, $\alpha=0^{\circ}$ (blue  dashed line) and  $\alpha=45^{\circ}$ (green solid line). (b) Time dependence of the fidelity for the atom coupled to Fock state field. Other parameters used are  $n=0$ and $\Delta=5$ Hz.}
\end{figure}

The estimation of the detuning is investigated by coupling the atoms to a single mode cavity field in this paper. In Fig.~1(a), the QFI of the detuning is plotted when the initial state of the system is prepared in the separable or entangled state. Figure 1(a) shows the QFI has oscillatory and rising behaviors in any initial state of the system, which means a better precision of the detuning estimation can be got by controlling interaction time. Additionally, the information of the detuning flows from probe system to the environment, that is, the non-Markovianity appears in the periods. This indicates the existence of the memory effect. We find that  the maximal QFI of the detuning estimation is $1.14\times 10^{3}$ when the initial state of the system is a separable state, while the maximal QFI is $1.18\times 10^{3}$ when the initial state of the system is a entangled state. The best precision of the detuning estimation is obtained via substituting the value of QFI into Cram\'{e}r-Rao inequality with $\nu=1$. Although, the entangled state is more advantageous to improve the precision of the detuning estimation, its result is very close to the result of the separable state. Time dependence of the fidelity for the atom coupled to Fock state field is shown in Fig.~1(b). More interestingly, Fig.~1(b) also presents an oscillatory behavior and the value of the fidelity is higher when the initial state of the system is the entangled sate. From Figs.~1(a) and (b), one can conclude that the entangled state is more beneficial to estimate the detuning. On the other hand, when the initial state is separable, the maxima of fidelity corresponds to the best precision of the detuning estimation.

\begin{figure}{}
	\centering
	\subfigure[]{
		\includegraphics[width=6cm]{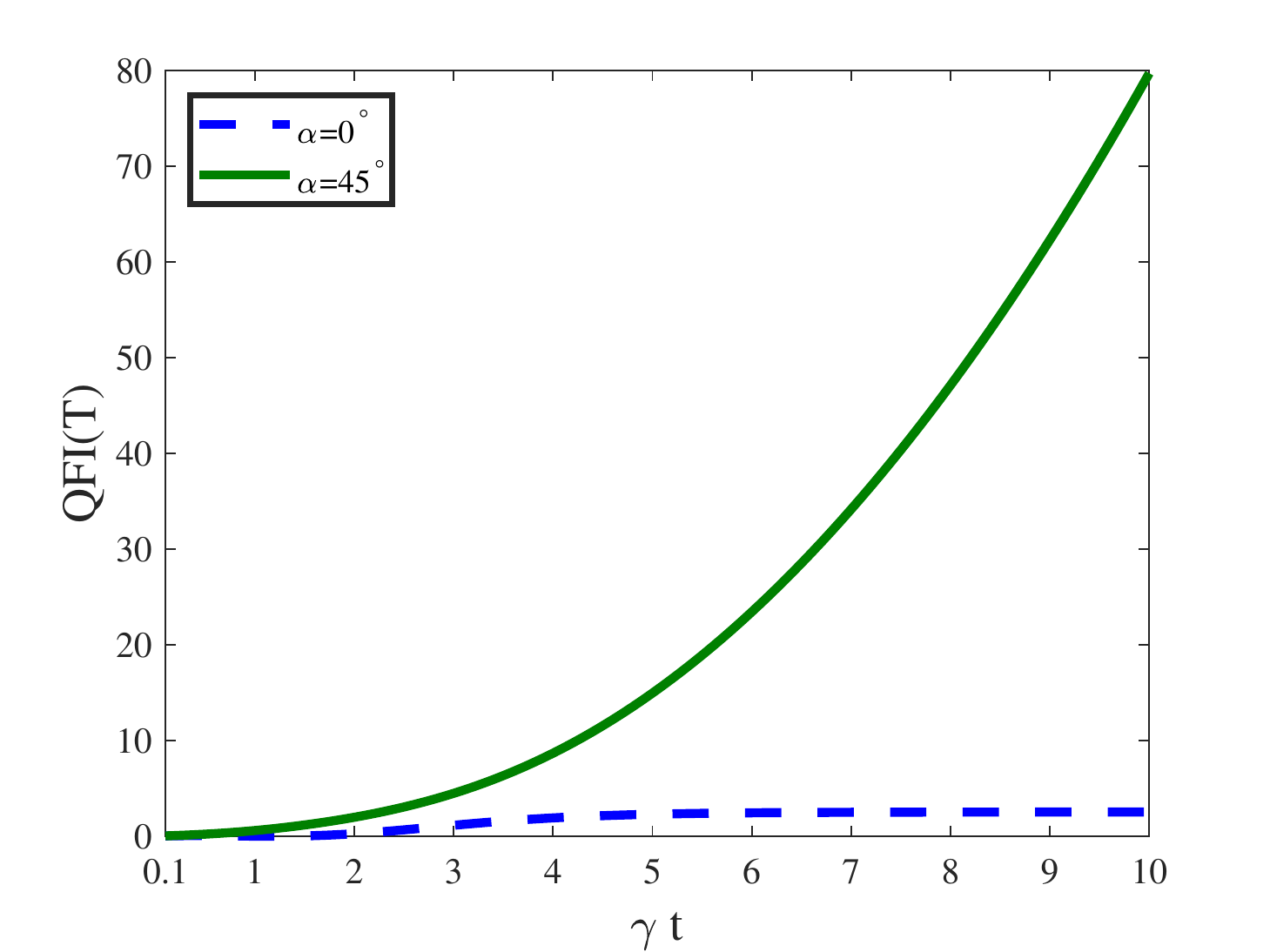}}
	\subfigure[]{
		\includegraphics[width=6cm]{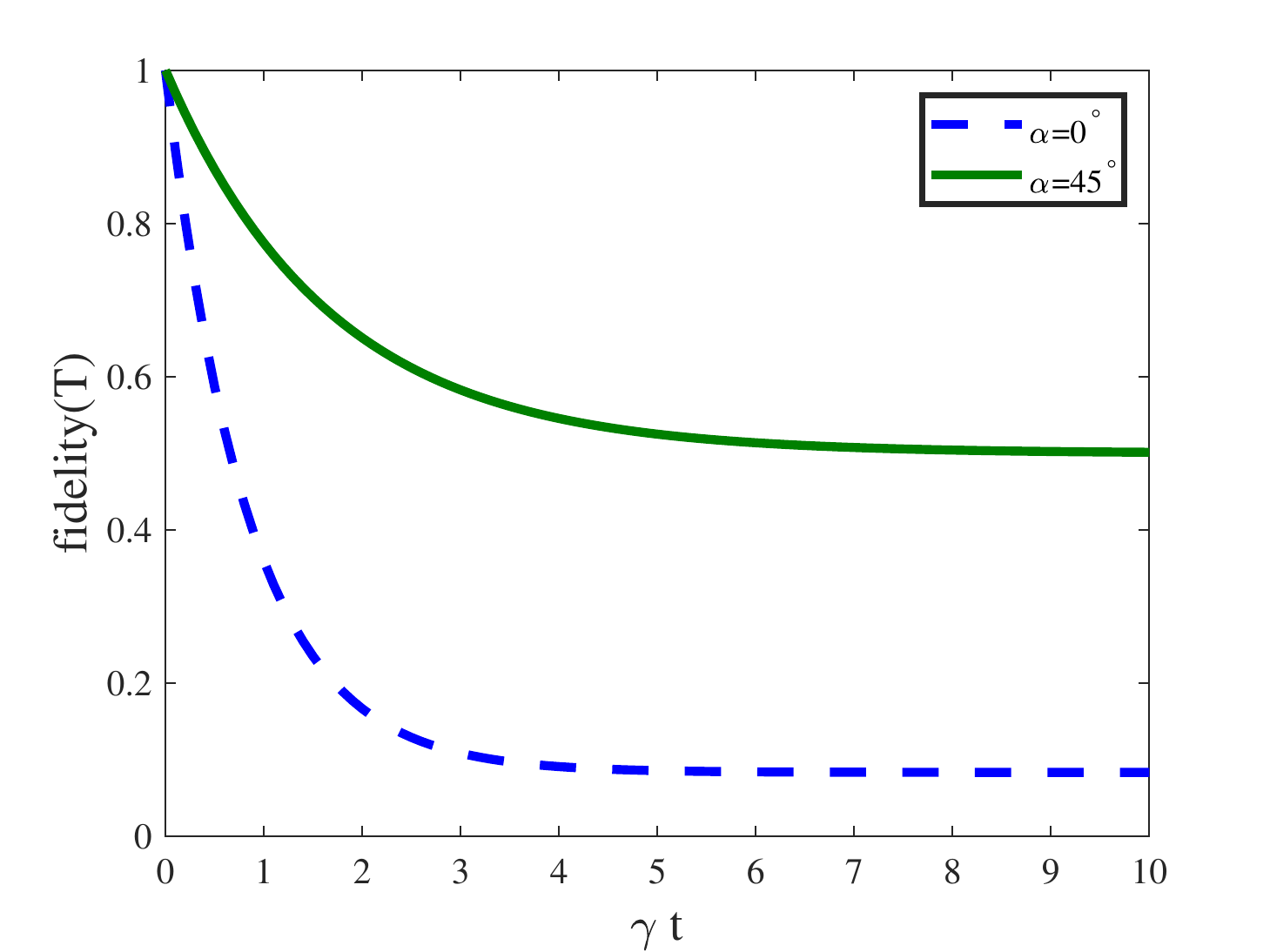}}
	\caption{\label{fig_double} (Color online) (a) QFI of the temperature $T$ as a function of time for different $\alpha$, $\alpha=0^{\circ}$ (blue dashed line) and  $\alpha=45^{\circ}$ (green solid line). (b) Time dependence of the fidelity for the atom interacts with a thermal reservoir. Other parameters used are $\frac{\hbar \nu_{k}}{k_{B}}=1$, $m=0.1$ and $\gamma=1$.}
\end{figure}

The QFI of the temperature is presented in Fig.~2(a) for the different initial atomic states. As demonstrated in Fig.~2(a), the QFI  of the temperature is enhanced with the increase of the time when the initial atomic state is the optimal superposition state, while the QFI grows quickly at first and then it isn't changed over time when the initial atomic state is non-superposition. Time dependence of the fidelity of the atom coupled to a thermal reservoir is plotted in Fig.~2(b) which indicates the fidelity is bigger when initial atomic state is the optimal superposition state. Therefore, the optimal superposition state is the best choice for estimating the temperature with a high precision. And the maximal QFI of the temperature is $80$ with one-qubit probe.

\begin{figure}{}
	\centering
	\subfigure[]{
		\includegraphics[width=6cm]{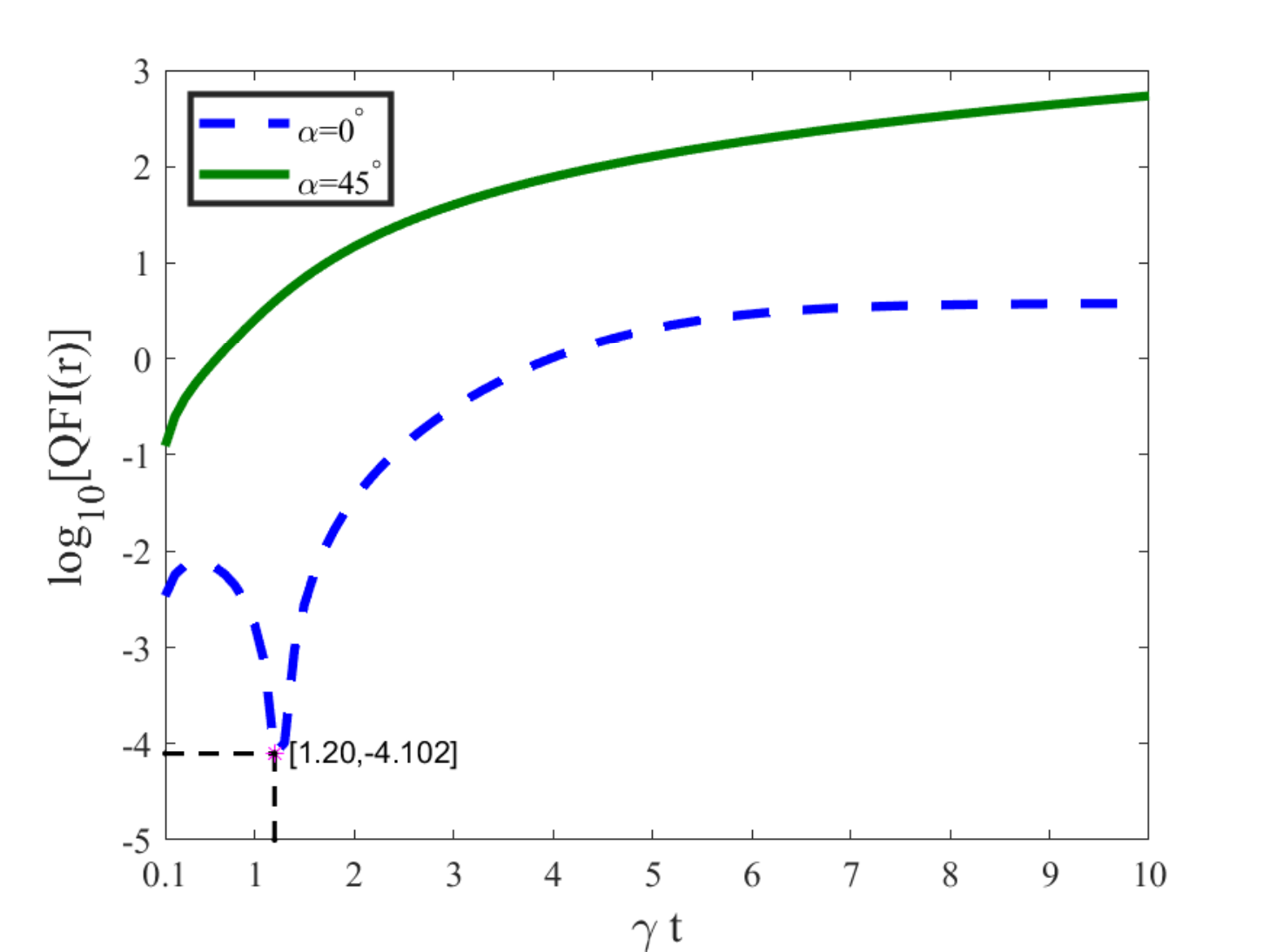}}
	\subfigure[]{
		\includegraphics[width=6cm]{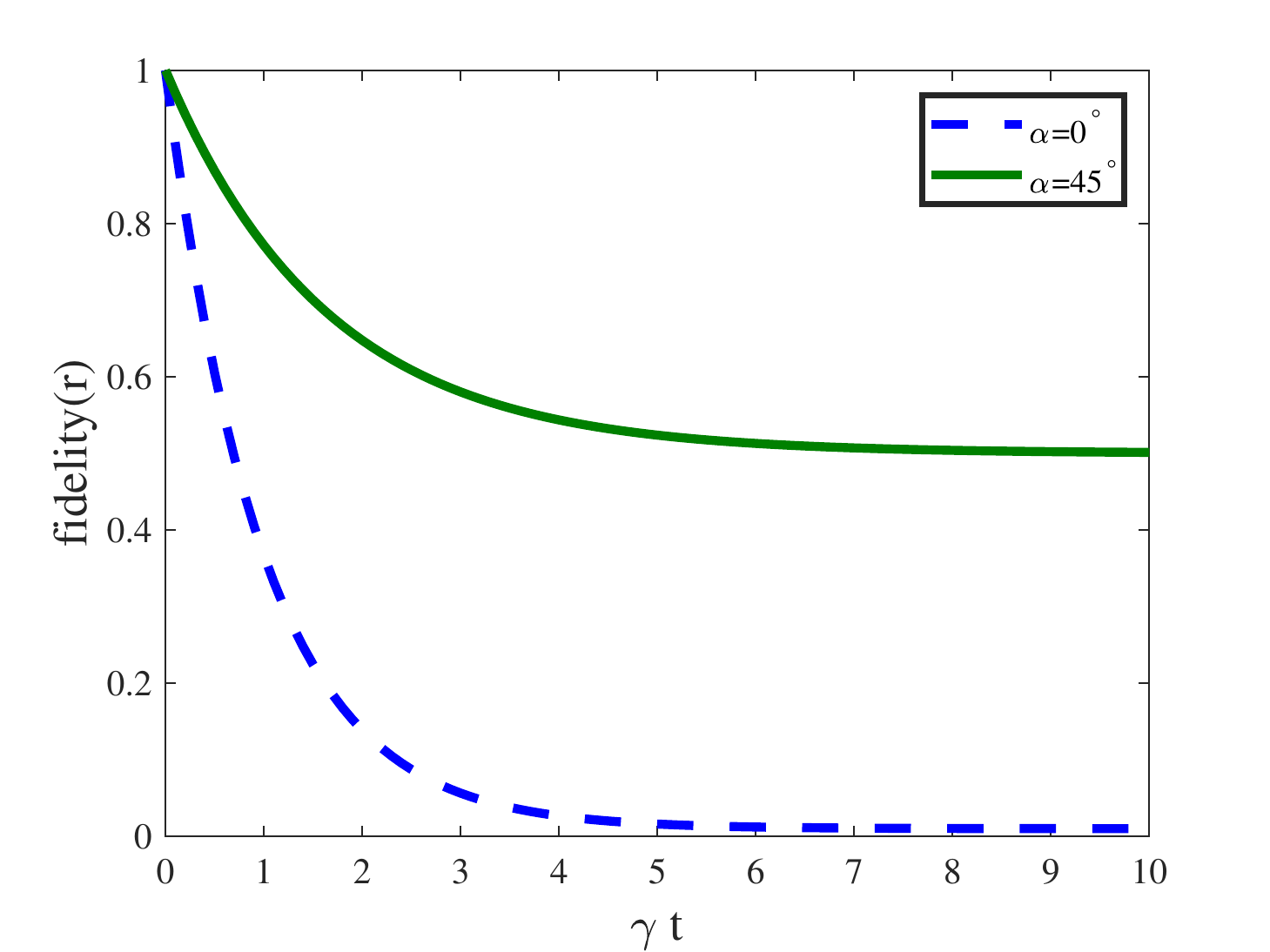}}
	\caption{\label{fig_double} (Color online) (a) QFI of the squeezing strength $r$ as a function of time for different $\alpha$, $\alpha=0^{\circ}$ (blue dashed line) and  $\alpha=45^{\circ}$ (green solid line). (b) Time dependence of the fidelity for the atom interacts with a squeezed  vacuum reservoir. Other parameters used are $\theta=0$, $r=0.1$ and $\gamma=1$.}
\end{figure}

The last parameter estimated is the squeezing strength. Fig.~3(a) shows the QFI of the squeezing strength as the function of dimensionless time $\gamma t$. From Fig.~3(a), one can find the increase of the interaction time is beneficial to estimate the squeezing strength. When $\alpha=0^{\circ}$, there is the presence of the QFI revivals at initial instants. It is caused by the followings, one is that the process of the information of the squeezing strength is encoded into the atom, the other is the decoherence effects. Time dependence of the fidelity for the atom immersed into a squeezed vacuum reservoir is exhibited in Fig.~3(b). Fig.~3(b) shows the final atomic state is completely different from the initial atomic state when $\alpha=0^{\circ}$. Comparing Figs.~3(a) and 3(b), we find that optimal superposition state can improve the estimation precision of the squeezing strength. The estimation of the squeezing strength shows the similarity with the case of the temperature estimation. The maximal QFI of the the squeezing strength is $5.41\times10^{2}$ with two-qubit probe.

\begin{figure}{}
	\centering
	\subfigure[]{
		\includegraphics[width=6cm]{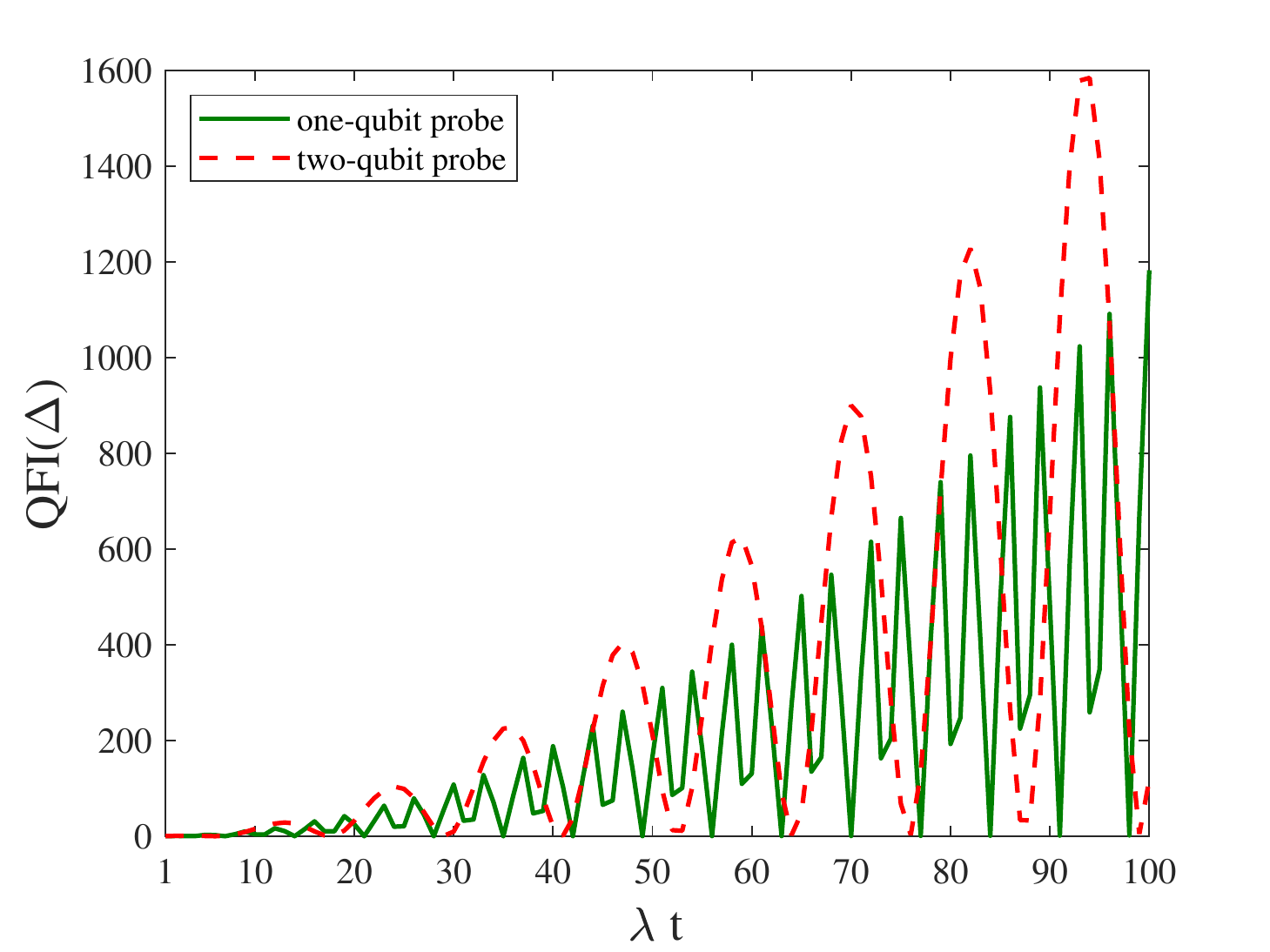}}
	\subfigure[]{
		\includegraphics[width=6cm]{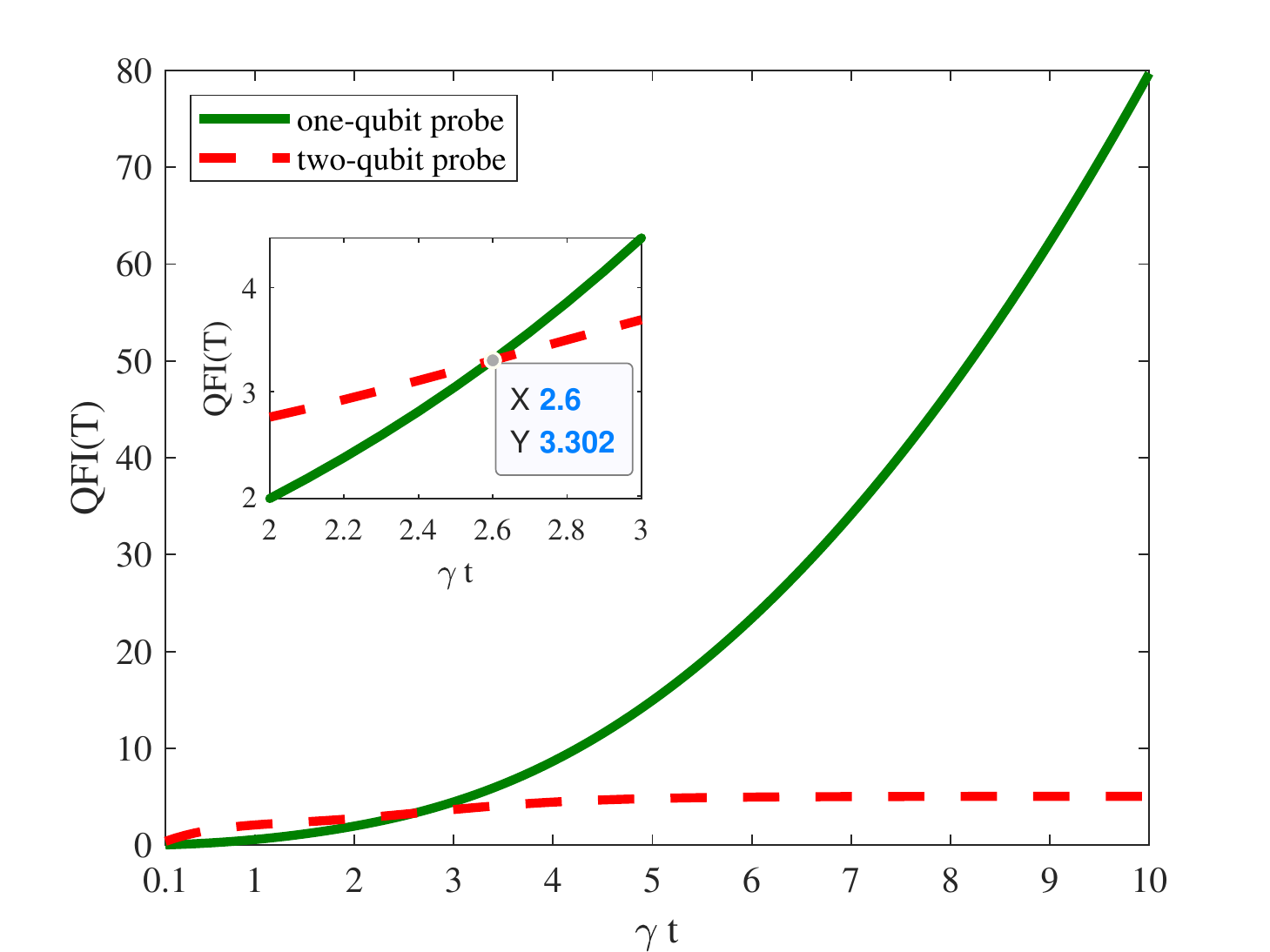}}
	\subfigure[]{
		\includegraphics[width=6cm]{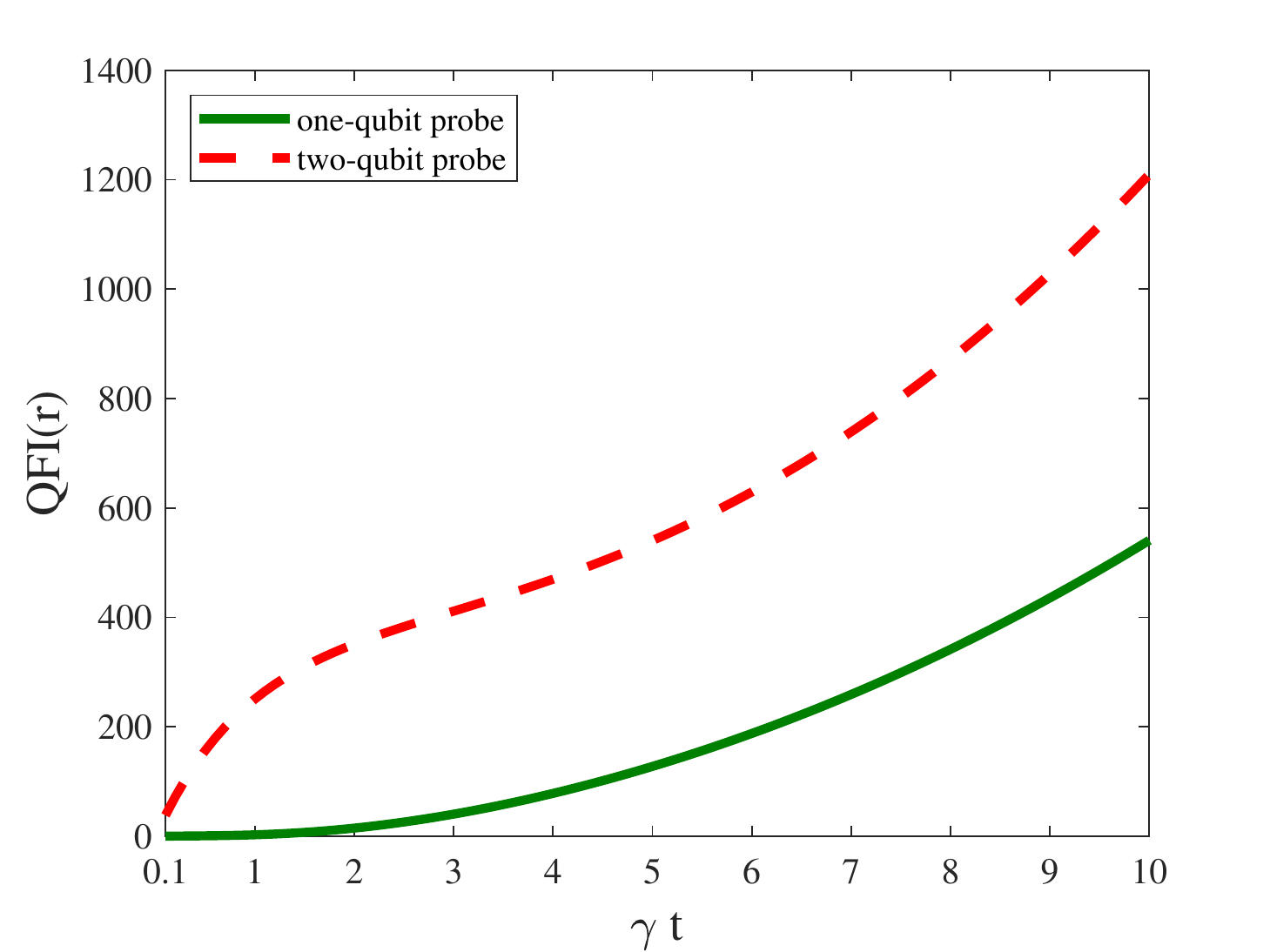}}
	\caption{\label{fig_double} (Color online) The QFI of (a) the detuning, (b) temperature and (c) the squeezing strength with two-qubit probe. The green lines correspond to the one-qubit probe. The red dashed lines denote the two-qubit probe. The other parameter used as Figs.~1(a), 2(a) and 3(a), respectively.}
\end{figure}

\begin{figure}{}
	\centering
	\subfigure[]{
		\includegraphics[width=6cm]{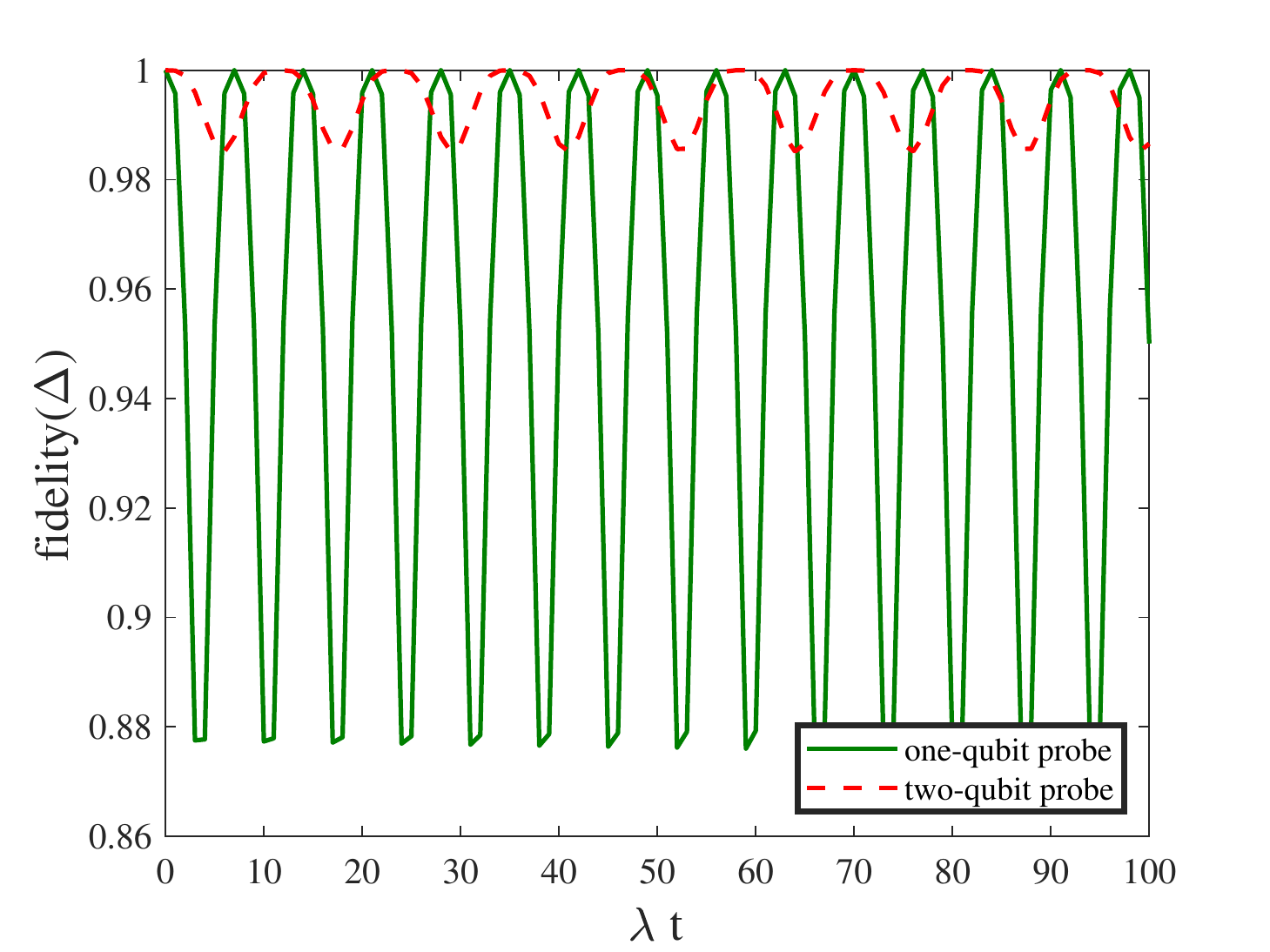}}
	\subfigure[]{
		\includegraphics[width=6cm]{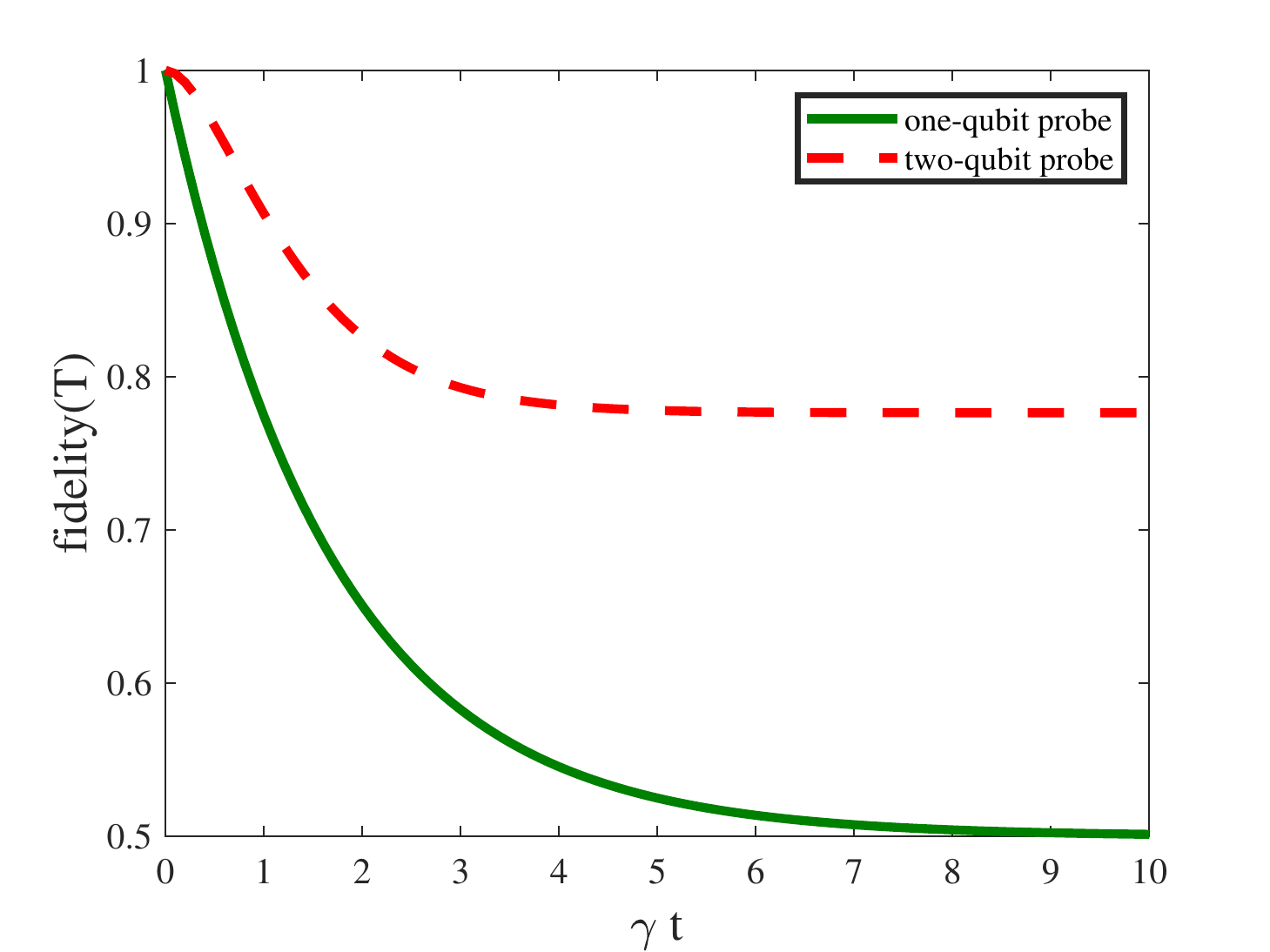}}
	\subfigure[]{
		\includegraphics[width=6cm]{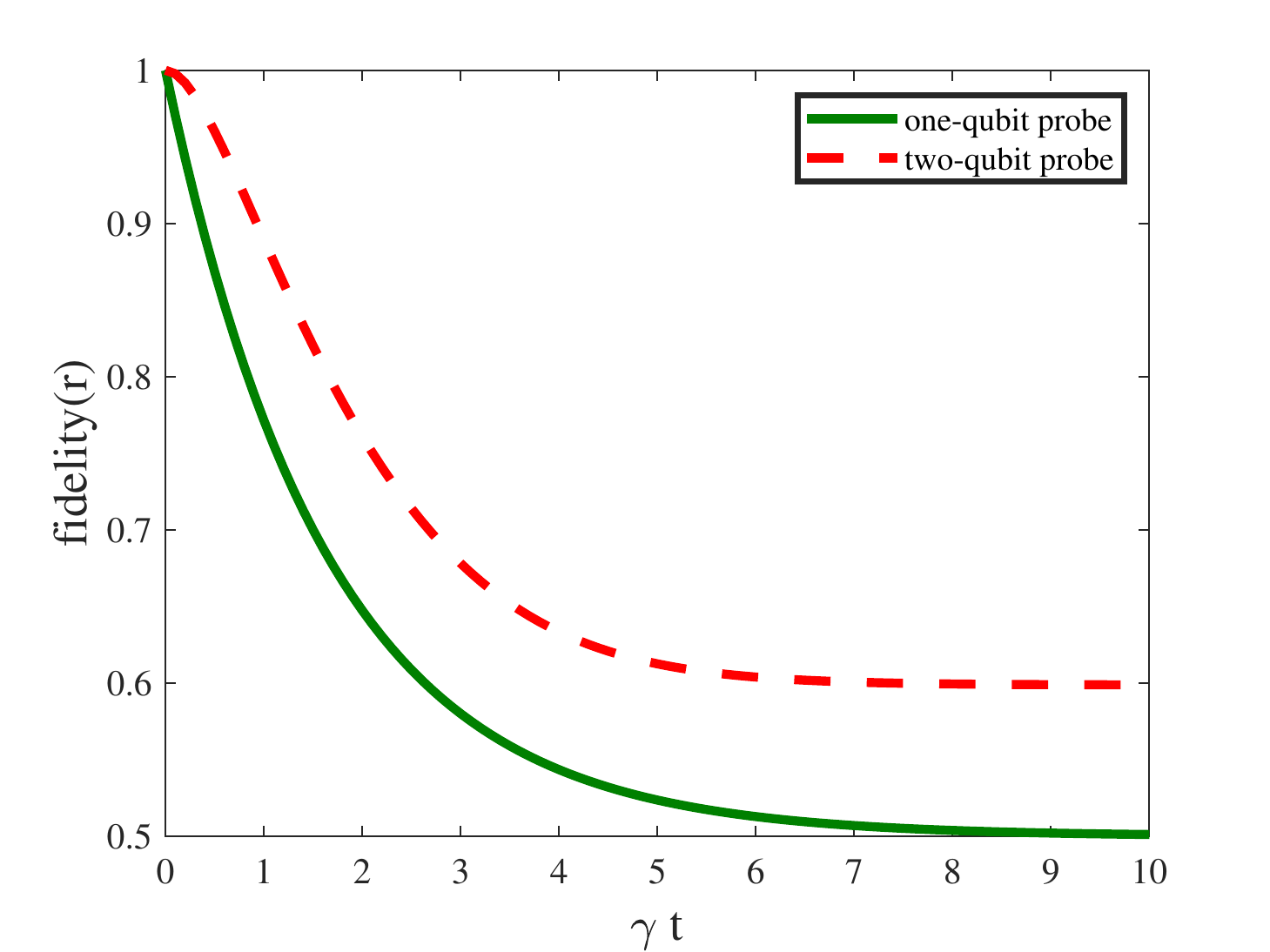}}
	\caption{\label{fig_double} (Color online) Time dependence of the fidelity of the atom A when the atoms interacts with (a) the fock state field, (b) the thermal reservoir and (c) the squeeze vacuum reservoir. The green lines correspond to the one-qubit probe. The red dashed lines denote the two-qubit probe. The other parameter used as Figs.~1(b), 2(b) and 3(b), respectively.}
\end{figure}

The QFI dynamics of two two-level atoms is drawn in Fig.~4. Surprisingly, the QFI of the detuning appears the behavior of collapse and revival, which manifests that there is the memory effect. The estimation precision of the detuning is enhanced with increasing time by ignoring the oscillation phenomena in Fig. 4(a). The maximal QFI shown in  Fig. 4(a) is $1.58\times10^{3}$ for the detuning estimation with two-qubit probe when $\gamma t=94$. The maximal QFI of the temperature is approximately $5$ in Fig. 4(b), which is far less than the one-qubit probe with the optimal superposition state as the initial atomic state. Fig. 4(c) shows the maximal QFI of the squeezing strength is $1.21\times10^{3}$ with two-qubit probe.

The fidelity of the atom A is plotted in Fig.~5 when the atoms interact with the Fock state field, the termal reservoir and the squeezed vacuum reservoir, respectively. In fig.~5, the green lines, which have been shown in Figs.~(1b), (2b) and (3b), respectively, correspond to the one-qubit probe, while the red dashed lines correspond to the two-qubit probe. Figure 5 indicates that the fidelity of the atom A is enhanced with the two-qubit probe. And the maximum value of QFI is obtained when the fidelity maxima is reached via two-qubit probe.

\section{Conclusions}
\label{sec:Conclusions}
In this paper, a novel approach was presented to estimate the system parameter, the detuning, and the enviroment parameter, temperature and the squeezing strength. We concluded that entangled state and optimal superposition state could improve the estimation precision of parameters with one-qubit probe via QFI and fidelity. For the estimation of the detuning, temperature and the squeezing strength, the fidelity of the atom is obviously enhanced via two-qubit probe compared with the one-qubit probe. The maximal QFI of the above estimated parameters are  $1.58\times10^{3}$, $80$ and $1.21\times10^{3}$, respectively, Based on the the quantum Cram\'{e}r-Rao bound, the estimation precision of parameters can be obtained. And the precision of estimated parameters can be further improved via increasing the interaction time. When the detuning between the cavity and the two-level atoms is estimated, there is the memory effect. The estimated parameter and the initial state of the probe system play an important role in the presence the QFI revivals when the atoms interact with the reservoir. The result of this paper has a potential application in absolute distance measurement, laser spectrum detection, atomic clocks, quantum thermometry, quantum communication and quantum information processing. One possible future goal is to optimize this model for estimating the parameters with a high precision in a shot interaction time. Another goal is to consider the environmental influence on estimating the detuning.

\section*{Acknowledgments}
This work was supported by the National Natural Science Foundation of China (Grant No. 91536115 and No. 11534008); Natural Science Foundation of Shaanxi
Province (Grant No. 2016JM1005).

\end{document}